\documentclass[journal,twoside, web]{ieeecolor}
\usepackage{generic}
\usepackage{cite}
\usepackage{amsmath,amssymb,amsfonts}
\usepackage{algorithmic}
\usepackage{graphicx}
\usepackage{textcomp}
\usepackage{siunitx}
\def\BibTeX{{\rm B\kern-.05em{\sc i\kern-.025em b}\kern-.08em
    T\kern-.1667em\lower.7ex\hbox{E}\kern-.125emX}}
\begin{document}

\title{ScAlN-on-SiC K$_u$-Band Sezawa Solidly-Mounted Bidimensional Mode Resonators}
\author{Luca Colombo, \IEEEmembership{Member, IEEE}, Luca Spagnuolo, \IEEEmembership{Student Member, IEEE}, Kapil Saha, Gabriel Giribaldi, \IEEEmembership{Student Member, IEEE}, Pietro Simeoni, \IEEEmembership{Member, IEEE}, and Matteo Rinaldi, \IEEEmembership{Senior Member, IEEE}
\thanks{Manuscript received November 16, 2024; revised December 29, 2024; accepted \textcolor{red}{add date}. Date of publication \textcolor{red}{add date}; date of current version \textcolor{red}{add date}. This work was supported by the Defense Advanced Research Project Agency Microsystems Technology Office (DARPA MTO) COFFEE TA1 project under Contract HR001122C0088 and developed in collaboration with RTX Missiles and Defense and Naval Research Lab (NRL). The review of this letter was arranged by Editor \textcolor{red}{add name}. (Corresponding authors: L. Colombo and M. Rinaldi.)}
\thanks{All the authors are affiliated with the Department of Electrical and Computer Engineering and the Institute for NanoSystems Innovation (NanoSI), Northeastern University, Boston, MA 02115 USA (e-mails: l.colombo@northeastern.edu, m.rinaldi@northeastern.edu).}
\thanks{L. Colombo and L. Spagnuolo contributed equally to this work. The authors would also like to thank Northeastern University
Kostas Cleanroom and Harvard Center for Nanoscale Systems (CNS) staff.}
\thanks{Color versions of one or more figures in this letter are available at \textcolor{red}{link}.}
\thanks{Digital Object Identifier \textcolor{red}{link}.}}
\maketitle

\begin{abstract}
This letter reports on Solidly-Mounted Bidimensional Mode Resonators (S2MRs) exploiting a highly-optimized Sezawa mode in 30\% Scandium-doped Aluminum Nitride (ScAlN) on Silicon Carbide (SiC) and operating near 16 GHz. Experimental results demonstrate mechanical quality factors ($Q_m$) as high as 380, Bode quality factors ($Q_{Bode}$) approaching 500, electromechanical coupling coefficients ($k_t^2$) of 4.5\%, an overall Figure of Merit ($FOM = Q_m \cdot k_t^2$) exceeding 17, and power handling greater than 20 dBm for devices closely matched to 50~$\Omega$. To the best of the authors' knowledge, S2MRs exhibit the highest Key Performance Indicators (KPIs) among solidly mounted resonators in the K$_u$-band, paving the way for the integration of nanoacoustic devices on fast substrates with high-power electronics, tailored for military and harsh-environment applications.

\end{abstract}

\begin{IEEEkeywords}

ScAlN, RF NEMS, Sezawa Mode

\end{IEEEkeywords}

\section{Introduction}
\label{sec:introduction}

Emerging communication paradigms are driving the demand for low Size, Weight, and Power (SWaP) Radio Frequency (RF) electronic components capable of performing signal processing, timing, and impedance matching in the K$_u$ band and beyond \cite{dunworth_28ghz_2018}\cite{zhong_-chip_2017}. Specifically, in the context of electromagnetic (EM) spectrum filtering, advanced 5G and proposed 6G commercial applications stand to benefit significantly from the development of compact, high-performance, and mass-manufacturable on-chip filters designed for handheld and wearable devices \cite{aigner_baw_2018}\cite{matthaiou_road_2021}. Similarly, Satellite Communications (SATCOM) and military applications, such as Active Electronically Scanned Arrays (AESA) \cite{gultepe_1024-element_2021}, would greatly benefit from technologies that enable monolithic integration, high power handling, and reliable operation at elevated temperatures (up to hundreds of degrees Celsius) \cite{alexandru_monolithic_2013}\cite{yuvaraja_wide_2023}.

While traditional EM filters \cite{zheng_compact_2017}\cite{xu_synthesis_2016} and novel spintronic devices \cite{du_frequency_2024} remain bulky and challenging to scale, researchers have increasingly turned their attention to piezoelectric acoustic Micro- and Nanoelectromechanical Systems (M/NEMS) as a promising alternative for RF filtering at higher frequencies \cite{kochhar_x-band_2023}. Acoustic MEMS devices, such as Surface Acoustic Wave (SAW) \cite{takai_high-performance_2019} \cite{kimura} \cite{LN} and Film Bulk Acoustic Resonators (FBARs) \cite{ruby_decade_2011}\cite{hagelauer_microwave_2023}, represent the state-of-the-art solutions in the field, offering high selectivity and low insertion loss ($IL$) \cite{vetury_xbaw_2022}. However, their performance beyond 5 GHz has been hindered by challenges such as declining quality factors ($Q$) and the need to operate within the nanoacoustic domain, where sound propagation speeds are significantly lower than those of EM waves, increasing fabrication complexity \cite{kramer_57_2022}.

Recent advances have demonstrated that acoustic NEMS devices can retain high $Q$ factors even in the mmWave band \cite{giribaldi_up-scaling_2024}, making them viable candidates for integrated nanoacoustic filters with performance metrics compatible with modern communication systems stringent requirements \cite{giribaldi_compact_2024}.

\begin{figure}[!b]
\centerline{\includegraphics[width=\columnwidth]{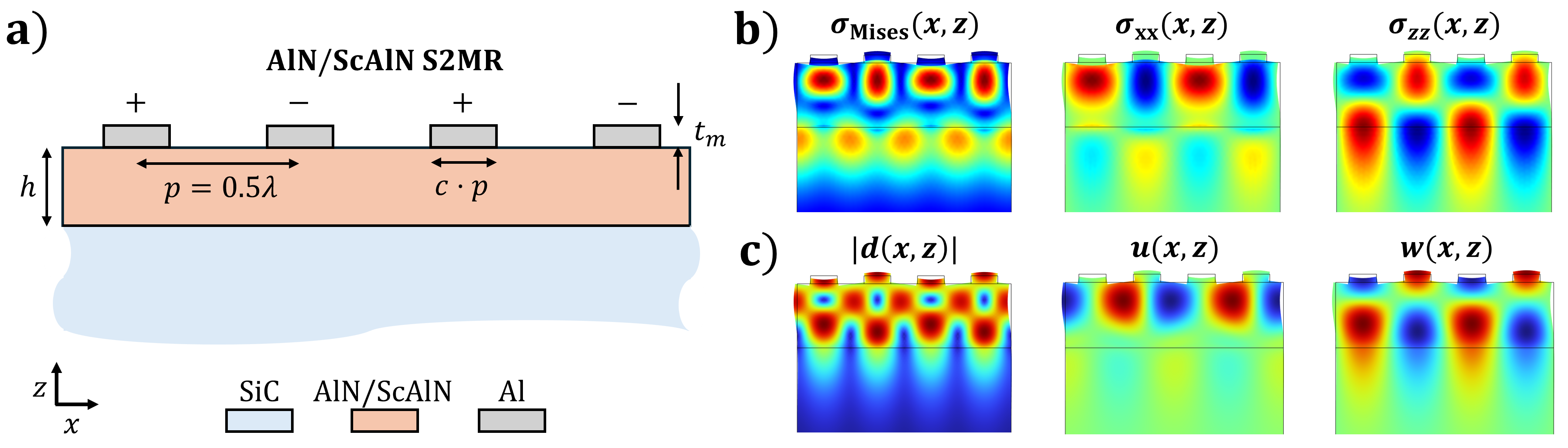}}
    \caption{a) Cross-section of the proposed slow-on-fast solidly mounted bidimensional mode resonator (S2MR). The S2MR is implemented on thin film  Scandium-doped Aluminum Nitride (ScAlN) on bulk 6H Silicon Carbide (SiC); 
    b) Von Mises equivalent stress ($\sigma_{Mises}$ and $x-$ and $z-$ stress tensors ($\sigma_{xx}$ and $\sigma_{zz}$, respectively) according to COMSOL\textregistered~Finite Element Analysis (FEA) simulations; c) Displacement magnitude ($d$) and displacements in the $x-$ and $z-$ directions ($u$ and $w$, respectively) according to COMSOL\textregistered~FEA simulations. Periodic mechanical and electrical boundary conditions (BCs) are applied to the lateral edges of the structure, while free BCs are selected for the ScAlN top and SiC bottom surfaces.}
    \label{fig:enter-label}
\end{figure}

In this letter, nanoacoustic Solidly-Mounted Bidimensional Mode Resonators (S2MR) operating at approximately 16 GHz are presented. This class of devices leverages a slow-on-fast optimized Sezawa mode in a thin film of 30\%-doped Scandium Aluminum Nitride (ScAlN) on Silicon Carbide (SiC) \cite{du_near_2024}. Key performance metrics include a mechanical quality factor ($Q_m$) of 380, an electromechanical coupling coefficient ($k_t^2$) greater than 4.5\%, and a resulting Figure of Merit (FoM) of 17. Additionally, S2MRs demonstrate power handling exceeding 20 dBm and offer a clear path for further performance enhancement. Their compatibility with monolithic integration on high-power electronic platforms such as SiC or diamond, combined with their robustness in harsh environments, positions these devices as a promising solution for high-performance filtering applications in both commercial and military domains. 

\section{Modeling}
\label{sec:modeling}

\begin{figure}[!t]
\centerline{\includegraphics[width=\columnwidth]{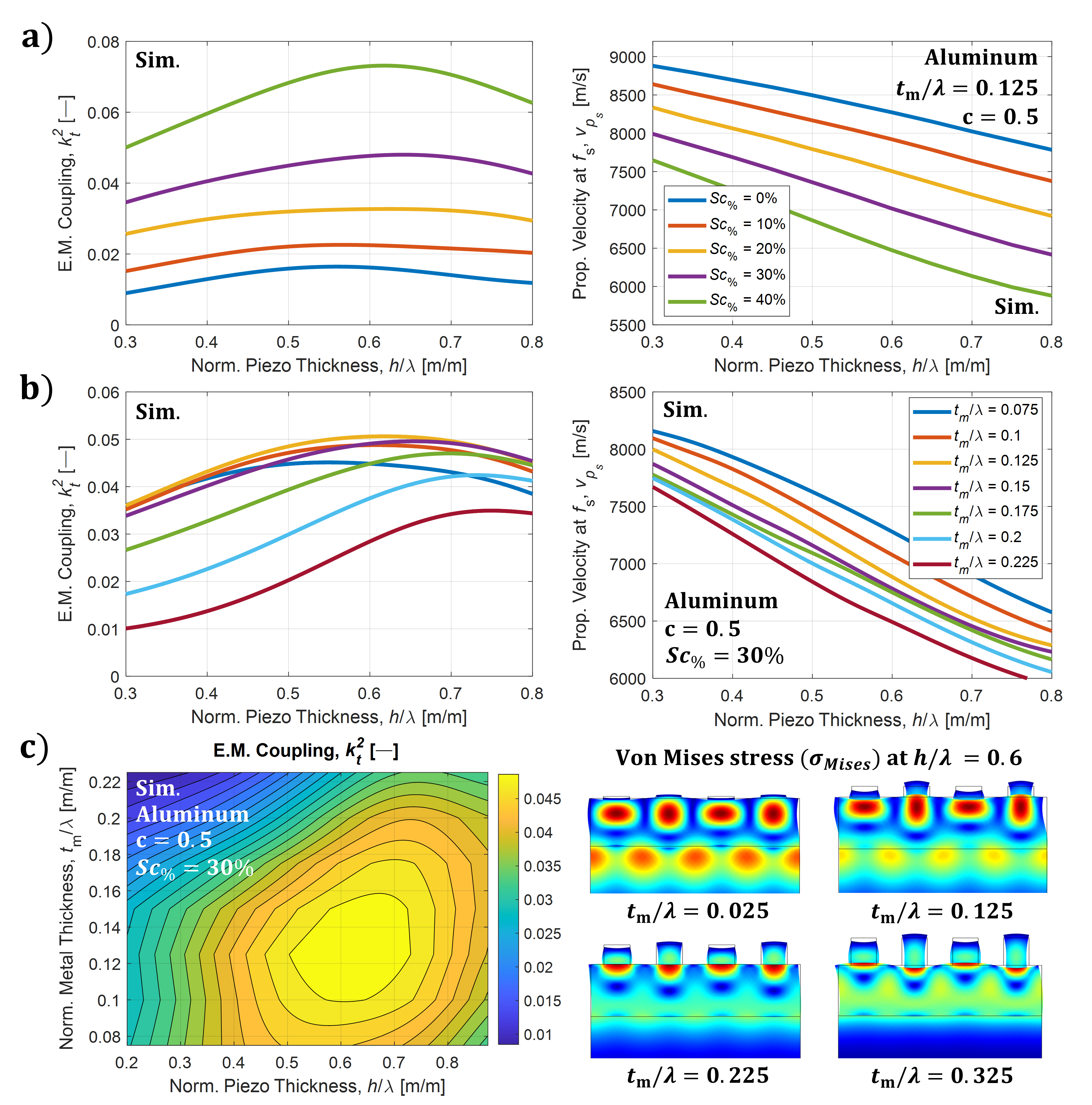}}
    \caption{a) COMSOL\textregistered~FEA simulated electromechanical coupling ($k_t^2$) and propagation velocity at resonance ($v_{p_s}$) as a function of normalized piezoelectric film thickness ($h$) over acoustic wavelength ($\lambda$) for Scandium-doping concentrations ($Sc_{\%}$) ranging between 0 and 40\%. Aluminum (Al) top electrodes with a coverage ($c$) of 50\% and normalized metal thickness ($t_m$) over $\lambda$ of 0.125 are assumed. Higher Sc concentrations soften the material, hence increasing the coupling and reducing the propagation velocity; b) $k_t^2$ and $v_{p_s}$ as a function of $h/\lambda$ for normalized Al $t_m/\lambda$ ranging between 0.075 and 0.225, $Sc_{\%}$ of 30\%, and 50\% IDT coverage; c) Contour plot of $k_t^2$ as a function of $h/\lambda$ and $t_m/\lambda$ for 50\% coverage Al IDTs and 30\% Sc-doping. A unique local maximum for $h/\lambda$ = 0.65 and $t_m/\lambda$ = 0.125 exists. FEA simulated Von Mises stress distributions for $h/\lambda$ = 0.6 and $t_m/\lambda$ ranging between 0.025 and 0.325, highlighting the electrode-induced stress shift towards the piezoelectric-metal interface when thicker IDTs are adopted. A $t_m/\lambda$ ratio of 0.125 ensure maximum coupling between the electric field and the stress while preserving most of the elastic strain energy within the piezoelectric layer.
    }
    \label{fig:optmize}
\end{figure}

The embodiment of the proposed Scandium Aluminum Nitride (ScAlN) nanoacoustic Solidly-Mounted Bidimensional Resonators (S2MR) consists of a thin film ($h$) of piezoelectric material solidly-mounted on a fast, bulk substrate of Silicon Carbide (SiC) with patterned metallic Interdigitated Transducers (IDTs) spaced by half acoustic wavelength ($\lambda$) on the top surface (Fig.~\ref{fig:enter-label}a). From a physical perspective, S2MRs are slow, dispersive Surface Acoustic Wave (SAW) devices exploiting a highly-optimized Sezawa mode \cite{ahmed_super-high-frequency_2023}\cite{hadj-larbi_sezawa_2019}\cite{sezawa_dispersion_1927} on a substrate with high propagation velocity ($v_p$). Similarly to Cross-Sectional Lamé Mode Resonators (CLMRs), they exploit both the $d_{33}$ and $d_{31}$ piezoelectric coefficient to generate a bidimensional acoustic standing wave with significant stress and displacement components in the lateral ($x$) and thickness ($z$) directions, as confirmed by COMSOL\textregistered~ Multiphysics Finite Element Analysis (FEA) simulations (Fig.~\ref{fig:enter-label}b-c). The large acoustic impedance mismatch between the active layer and the substrate is pivotal in constraining the acoustic energy into the piezoelectric medium, hence ensuring the existence of the resonant mode with large quality factor ($Q$) and electromechanical coupling ($k_t^2$). Metal-metal and metal-insulator layered Bragg reflectors \cite{barrera_18_2024}, sapphire \cite{guida_solidly_2023}, silicon carbide \cite{colombo_solidly_2024}, and diamond \cite{fujii_diamond_2020} substrates exhibit mechanical properties suitable for sustaining S2MRs in Sc-doped AlN.

Finite element analysis (FEA) simulations based on the \textit{a~priori} material properties of ScAlN, as described by Caro et al. \cite{caro_piezoelectric_2015}, reveal a super linear growth in $k_t^2$ at higher levels of Sc-doping ($Sc_{\%}$). This behavior is attributed to the combined effects of increased piezoelectric coefficients and a reduction in propagation velocity (Fig.~\ref{fig:optmize}a). Despite the presence of a local $k_t^2$ maxima for an $h/\lambda$ of 0.6, this class of devices retains sub-optimal, large coupling over a broad range of $v_p$, thus enabling single-step, lithographic multi-frequency definition for the synthesis of on-chip acoustic filter banks \cite{giribaldi_compact_2024}. Ultimately, S2MRs require tailored electrode material-thickness ($t_m$) optimization to maximize the coupling between the electric field laterally excited by the IDTs \cite{murata_improvement_2009} and the Sezawa's mode induced acoustic strain field. As reported in Figs.~\ref{fig:optmize}b-c, $h$, and $t_m$ must be co-optimized for a given IDT material, coverage ($c$), and $Sc_{\%}$, to ensure ideal energy localization in the piezoelectric medium.

In this work, 30\%-doped AlN-seeded ScAlN, 6H-SiC, and patterned aluminum-silicon-copper (AlSiCu) with a 50\% metallization ratio ($c$) are implemented as the active layer, fast substrate, and IDT metal, respectively. While the material selection is primarily driven by established fabrication processes \cite{giribaldi_compact_2024}, the choice of the fast substrate hints at the potential for future monolithic integration of RF acoustic front-ends with high-power electronics, tailored for military and harsh-environment applications. According to the FEA simulations reported in Fig.~\ref{fig:optmize}b, $h$ is set to 250~nm, $t_m$ to 50~nm, and $\lambda$ to 400~nm to target a frequency of operation ($f_s$) of 16 GHz and maximized coupling. 

\section{Fabrication}

\label{sec:devices}

\begin{figure}[!b]
\centerline{\includegraphics[width=\columnwidth]{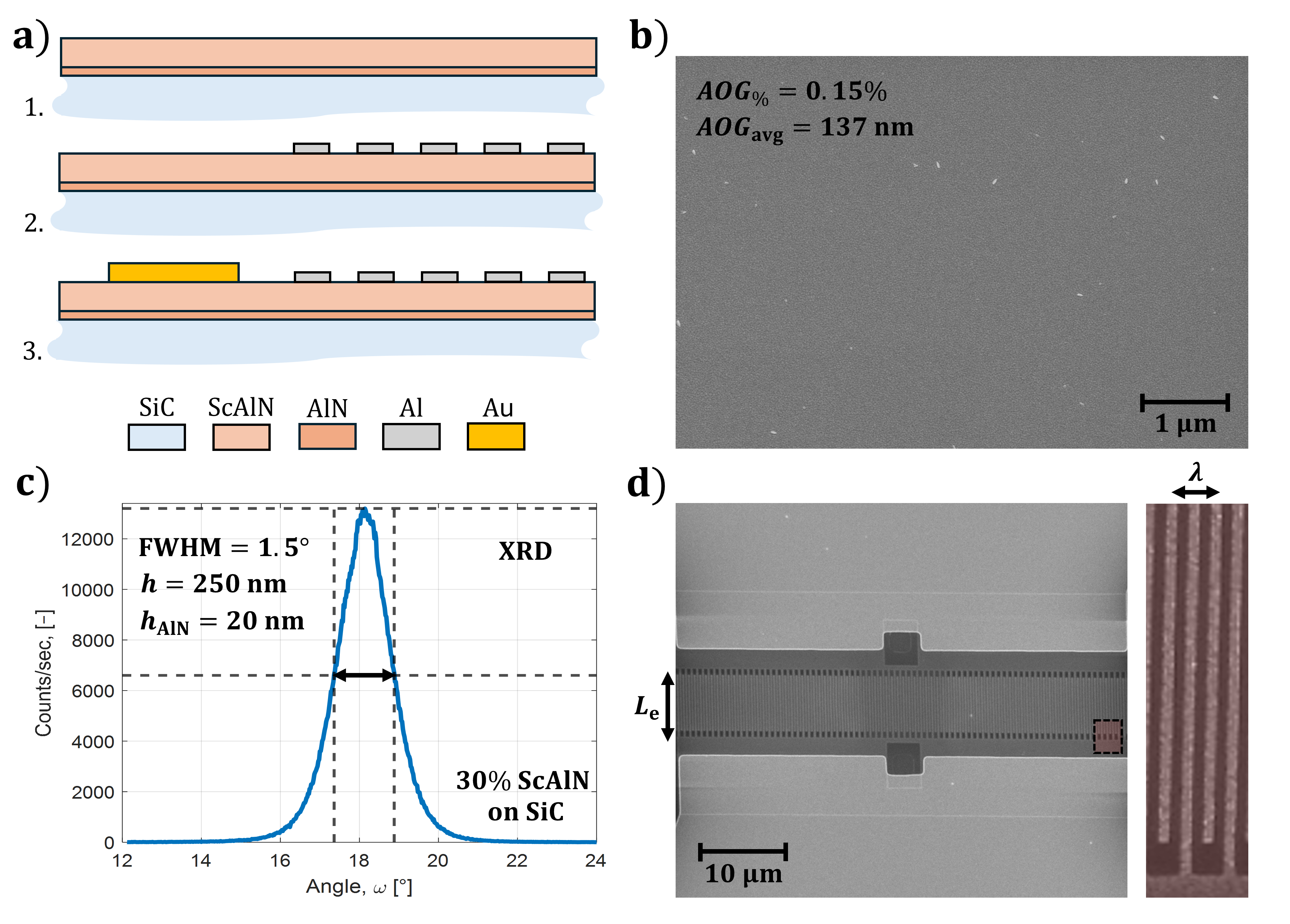}}
    \caption{a) Diagram of the micro-fabrication process adopted for the manufacturing of ScAlN S2MRs; 
    Top to bottom: 1) ScAlN reactive sputtering on top of a ultra-thin AlN seed layer; 2) Top electrode patterning via electron beam (e-beam) lithography, aluminum-silicon-copper (AlSiCu) thermal evaporation, and lift-off; 3) Pads and interconnects patterning via direct laser lithography, gold (Au) e-beam evaporation, and lift-off; 
    b) SEM image of the sputtered 30\%-doped ScAlN thin film surface, reporting low Abnormally Oriented Grains (AOGs) density; c) X-Ray Diffraction (XRD) Full-Width-Half-Maximum (FWHM) diagram, highlighting excellent crystallinity; and d) SEM image of a fabricated S2MR operating around 16 GHz. The reported device has an electrode aperture ($L_e$) of 5 \si{\micro\metre}, and acoustic wavelength ($\lambda$) of 400 nm, 120 finger pairs ($N_f$), and a finger width ($\lambda$/4) of 100 nm.}
    \label{fig:device}
\end{figure}

The fabrication process adopted for the S2MRs manufacturing is illustrated in Fig. \ref{fig:device}a. The process begins with the in-house deposition of a 230 nm layer of 30\% ScAlN onto a 4-inch SiC substrate via reactive sputtering, using a commercial-grade Evatec CLUSTERLINE\textregistered~200 II system and a 12-inch casted target. The ScAlN thin film is sputtered onto a pre-deposited 20 nm-thick AlN seed layer \cite{hardy_nucleation_2023}, which serves as a templated nucleation layer to promote the growth of highly crystalline, $c$-axis-oriented piezoelectric films \cite{felice_energetics_1996}. 

Characterization using Scanning Electron Microscope (SEM), Abnormally Oriented Grain (AOG) surface density ($AOG_{\%}$) and average dimension ($AOG_{avg}$) evaluation \cite{spagnuolo_image_2024}, as well as X-Ray Diffraction (XRD) Full Width Half Maximum (FWHM) measurements confirmed state-of-the-art crystallinity. The results show an almost AOG-free top surface ($AOG_{\%}$ = 0.15\% and $AOG_{avg}$ = 137~nm) and a Rocking Curve (RC) of 1.5\textdegree~(Figs.~\ref{fig:device}b-c). 

Subsequently, Interdigital Transducers (IDTs) are defined via e-beam lithography, necessitated by the critical feature size of 100 nm. A 50 nm thin film of aluminum-silicon-copper (AlSiCu) is deposited through thermal evaporation, followed by an overnight lift-off in 1165 Microposit remover. AlSiCu is adopted in place of pure aluminum due to its increased resistance to electromigration \cite{braunovic_electrical_2017}. Finally, pads and interconnects are patterned using thermal evaporation of 250 nm of gold (Au), followed by a lift-off process. An SEM image of a fabricated devices is reported in Fig.~\ref{fig:device}d.

\section{Results and Discussion}
\label{sec:meas}

\begin{figure}[!t]
\centerline{\includegraphics[width=\columnwidth]{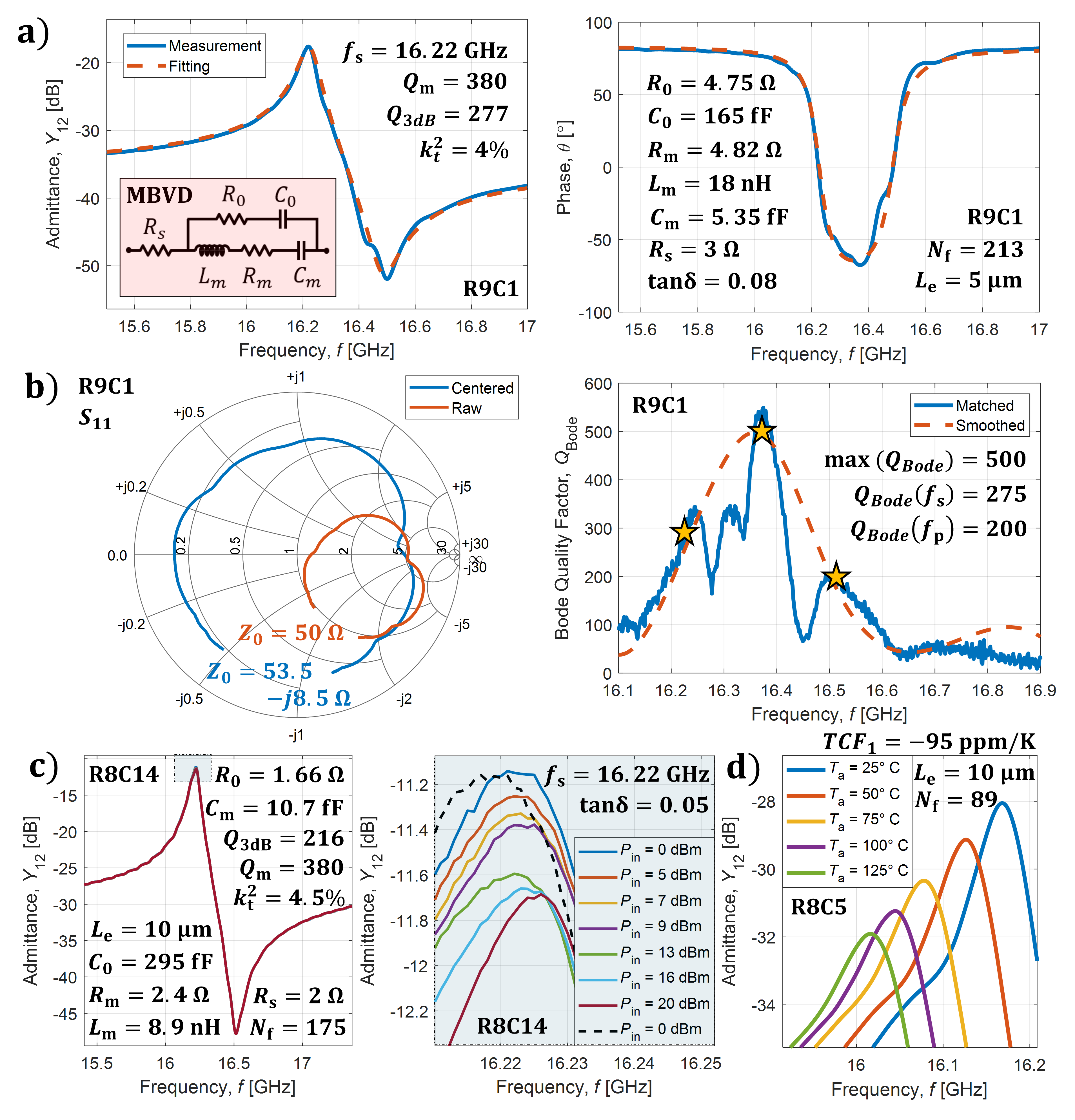}}
    \caption{a) Admittance response ($Y_{12}$) and phase ($\theta$) of a fabricated S2MR operating around 16 GHz and exhibiting the largest 3-dB quality factor ($Q_{3dB}$). The device is fit to a Modified Butterworth-Van Dyke (MBVD) model, which is reported in the inset. The 3-dB quality factor is calculated as $Q_{3dB}$ = $f_s/BW_{3dB}$, while the electromechanical coupling is calculated as $k_t^2 = \pi/2 \cdot f_s/f_p \cdot 1/\tan(\pi/2 \cdot f_s/f_p)$; b) Smith chart representation of the measured $S_{11}$ reflection coefficient parameter before (red line) and after (blue line) matching for Bode $Q$ extraction and inferred matched and smoothed $Q_{Bode}$ as a function of frequency;
    c) Admittance response ($Y_{12}$) and zoomed-in view of the peak region (blue box) of the S2MR showcasing the largest mechanical quality factor ($Q_m$) and electromechanical coupling ($k_t^2$) for different levels of applied input power ($P_{in}$). Motional parameters are reported for $P_{in}$~=~0~dBm. The legend details the tests as they were conducted, with the black dashed line indicating a return to baseline; and d) Temperature Coefficient of Frequency (TCF) testing for device R8C5 and applied temperature ($T_a$) ranging between 20\textdegree C and 125\textdegree C.
    }
    \label{fig:power}
\end{figure}

The fabricated devices are characterized under laboratory conditions ($T$ = 293 K) with 150 \si{\micro\meter} GSG probes in a two-port configuration to extract their characteristics frequency response. Scattering ($S-$) parameters are measured with a vector network analyzer (Keysight P5008A) and later converted into admittance ($Y-$) parameters via software. The equivalent circuit parameters are extracted by fitting the transmission admittance ($Y_{12}$) response to a Modified Butterworth-Van Dyke (MBVD) equivalent model \cite{larson_modified_2000}. 
The admittance of the device exhibiting the largest 3-dB quality factor ($Q_{3dB}$) is reported in Fig.~\ref{fig:power}a, demonstrating a $Q_{3dB}$ of 277, a motional quality factor of $Q_m$, and $k_t^2$ of 4\%, for an overall Figure of Merit ($FOM \approx Q_m \cdot k_t^2$) slightly exceeding 14. A Bode quality factor ($Q_{Bode}$) of 500 is extracted from the same device (Fig.~\ref{fig:power}b). Power handling characterization is performed on a separate device, most closely matched to 50~$\Omega$, possessing similar quality factor ($Q_m$ = 380 and $Q_{3dB}$ = 216) and the highest measured coupling ($k_t^2 = 4.5\%$) and Figure of Merit ($FOM$ = 17). Fig.~\ref{fig:power}c show the admittance response as the applied input power is swept from 0 dBm to the maximum power permitted by the setup (20~dBm), and then returned to the baseline of 0 dBm. A close-up of the resonant peak reveals minimal admittance degradation across the investigated input power range and no deterioration of the quality factor upon returning to baseline. A first-order Temperature Coefficient of Frequency ($TCF_1$) of -95 ppm/K is finally inferred from temperature testing ($T_a$) between 293 and 318~K (Fig.~\ref{fig:power}d). 

\section{Conclusion}
\label{sec:conclusion}

To the best of the authors’ knowledge, S2MRs exhibit the highest Key Performance Indicators (KPIs) among solidly mounted resonators in the K$_u$-band, paving the way for the monolithic integration of nanoacoustic devices with high-power electronics on SiC for commercial, military, and harsh-environment applications requiring sharp frequency selectivity and high power handling.

\clearpage
\newpage
\bibliographystyle{ieeetr} %
\bibliography{bibliography}

\end{document}